%% file: fqsq_L4Rv2.tex
\documentclass{elsart}
\usepackage[dvips]{graphicx}
\newcommand{\kmunudk}{\ensuremath{D^0 \rightarrow K^- \mu^+ \nu }}
\newcommand{\dstardk}{\ensuremath{D^{*+} \rightarrow D^0 \pi^+}}
\newcommand{\kpidk}{\ensuremath{D^0 \rightarrow K^- \pi^+}}
\newcommand{\mdifflong}{\ensuremath{\delta m \equiv m(K^- \mu^+ \nu~~ \pi^+) -m(K^- \mu^+ \nu)}}
\newcommand{\mdiff}{\ensuremath{\delta m}}
\newcommand{\mpole}{\ensuremath{m_{\rm pole}}}
\newcommand{\etak}{\ensuremath{f_-^{(K)}(0)/f_+^{(K)}(0)}}
\newcommand{\etakresult}{\ensuremath{\etak = -1.7^{+1.5}_{-1.4}\pm 0.3}}
\newcommand{\mpoleresultpi}{\ensuremath{\mpole{}= 1.91^{+0.30}_{-0.15} \pm 0.07~\gevcsq}}
\newcommand{\mpoleresultdecon}{\ensuremath{\mpole = 1.91 \pm 0.04 \pm 0.05~\gevcsq}}
\newcommand{\mpoleresult}{\ensuremath{\mpole = 1.93 \pm 0.05 \pm 0.03~\gevcsq}}
\newcommand{\alpharesult}{\ensuremath{\alpha = 0.28 \pm 0.08 \pm 0.07}}
\newcommand{\pimunudk}{\ensuremath{D^0 \rightarrow \pi^- \mu^+ \nu }}

\newcommand{\gevcsq}{\ensuremath{\textrm{GeV}/c^2}}

\newcommand{\thl}{\ensuremath{\theta_\ell}}

\newcommand{\costhl}{\ensuremath{\cos\thl}}

\newcommand{\qsq}{\ensuremath{q^2}}
\newcommand{\fplus}{\ensuremath{f_+(\qsq )}}
\newcommand{\fplusK}{\ensuremath{f_+^{(K)}(\qsq )}}
\newcommand{\fplusbin}{\ensuremath{f_+(q_i^2)}}
\newcommand{\fplusbinJ}{\ensuremath{f_+(q_j^2)}}
\newcommand{\fplusbinsq}{\ensuremath{f^2_+(q_i^2)}}
\newcommand{\fminus}{\ensuremath{f_-( \qsq )}}
\newcommand{\mdsstar}{\ensuremath{m_{D_s^*}}}
\newcommand{\mdpstar}{\ensuremath{m_{D^{+*}}}}
\newcommand{\mdstar}{\ensuremath{m_{D^*}}}

\newcommand{\gevc}{\ensuremath{\textrm{GeV}^2/c^2}}

\newcommand{\mysection}[1]{\section{#1}}

\newcounter{saveeqn}%

\begin{document}
\begin{frontmatter}
\title{Measurements of the \qsq{} dependence of the \kmunudk{} and \pimunudk{} form factors }
\input{authors_revised.tex}
\nobreak
\begin{abstract}
Using a large sample of \kmunudk{} and \pimunudk{} decays
collected by the FOCUS photoproduction experiment at Fermilab, we
present new measurements of the \qsq{} dependence for the 
\fplus{} form factor. These measured \fplus{} form factors are fit to common parameterizations such 
as the pole dominance form and compared to recent unquenched Lattice QCD calculations. We find \mpoleresult{} for \kmunudk{} and \mpoleresultpi{} for \pimunudk{} and \etakresult{}.
\end{abstract}
\end{frontmatter}
\newpage

\mysection{Introduction}

In this paper, we provide a new non-parametric measurement of the \qsq{}
evolution for the \fplusK{} form factor describing the pseudoscalar decay \kmunudk{}. The measurement is presented
in a form that is convenient for parametric and non-parametric comparisons
to other experiments and theoretical predictions. Our \qsq{} evolution is
compared to the lattice gauge calculations in \cite{simone}, and we show that fits
to the \qsq{} evolution agree with traditional parametric analyses of
the data and results from other experiments.  It is important\cite{cleo-c} to make
incisive tests of unquenched 
Lattice Gauge calculations in semileptonic decay processes such as \kmunudk{} and \pimunudk{}
in order to ultimately reduce the substantial systematic errors on the CKM matrix
in charm and related beauty processes.

Two form factors describe the matrix element for such decays according to Eq. \ref{ME}

\begin{eqnarray}
M = G_F V_{cs} \left[ \fplus~(P_D + P_K)_\sigma + \fminus~(P_D - P_K)_\sigma \right]~
{\bar u}_\mu \gamma^\sigma (1 - \gamma) u_\nu \phantom{xxxxx} 
\label{ME}
\end{eqnarray}
These lead to a differential width of the form given by Eq. \ref{dgamdq}
where $P_K$ is the kaon momentum in the $D^0$ rest frame and all \fminus{} contributions are multiplied by the square of the muon mass.\footnote{This form was obtained using the basic formulae in \cite{KS}.}

\begin{eqnarray}
{{d\Gamma}\over d \qsq}&=
{G_{F}^{2}\over{8\pi^{3}~m_D}}
\left|{V_{cs}}\right|^{2}
\left|{f_{+}(q^{2})}\right|^{2}
P_{K}\left({{W_{0}-E_{K}}\over F_0}\right)^2
\left[
{1\over 3}m_{D}P_{K}^2+
{m_{\mu}^{2}\over {8m_{D}}}(m_{D}^{2}+m_{K}^{2}+2m_{D}E_{K})
\right.\cr&\phantom{=G}\left.
+{1\over 3}m_{\mu}^{2}{P_{K}^{2}\over F_{0}}
+{1\over 4}m_{\mu}^{2}{{m_{D}^{2}-m_{K}^{2}}\over m_{D}}
Re\left({f_{-}(q^{2})\over f_{+}(q^{2})}\right)
+{1\over 4}m_{\mu}^{2}F_0
\left|{f_{-}(q^{2})\over f_{+}(q^{2})}\right|^2
\right]
\label{dgamdq}
\end{eqnarray}
In Eq. \ref{dgamdq}, $W_{0}=\left( m_{D}^{2}+m_{K}^{2}-m_{\mu}^{2} \right)/ (2m_D)$,
$F_{0}=W_{0}-E_{K}+ m_{\mu}^{2}/ (2m_{D})$ and $P_K$ , $E_K$ are the momenta
and energy of the kaon in the $D^0$ rest frame.
Assuming $\fminus /\fplus$ is of the order of unity as expected, the corrections due to \fminus{} are 
fairly small and, apart from the low \qsq{} region, \newline 
${d \Gamma / d q^2} = G^2_F~|V_{c q}|^2 ~P^3_K ~|\fplus|^2 / (24 \pi^3)$ is an excellent approximation. 

This paper provides new measurements of the \fplus{} form factors for \kmunudk{} and \pimunudk{} and of the ratio
$f_-(0)/f_+(0)$ for \kmunudk{}.\footnote{Throughout this paper, we will assume that $f_-(\qsq)/f_+(\qsq)$ is
essentially independent of \qsq.}
Our emphasis in this paper is on the \emph{shape} of the \qsq{} dependence rather than on its absolute
normalization. As a means of comparing our result to different parameterizations commonly used in the literature, 
we will fit our measurements of \fplus{} to two different parameterizations in use: the pole form given by Eq. \ref{pole} and the \emph{modified} pole form given
by Eq. \ref{modpole}.\footnote{In this form, the parameter $\alpha$ gives the deviation of \fplus{} from  spectroscopic pole dominance where $\mdstar = \mdsstar{} = 2.112 ~\gevcsq$ for \kmunudk{} and $\mdstar = \mdpstar = 2.010~\gevcsq$ for \pimunudk{}}    

\begin{eqnarray}
 \fplus = {f_+(0) \over 1 - \qsq/\mpole^2}
 \label{pole}
\end{eqnarray}

\begin{eqnarray}
 \fplus = {f_+(0) \over \left(1 - \qsq/\mdstar^2\right)\left(1 - \alpha~\qsq/\mdstar^2\right)}
 \label{modpole}
\end{eqnarray}

Throughout this paper, unless explicitly stated otherwise,
the charge conjugate is also implied when a decay mode of a specific
charge is stated.

\mysection{Experimental and analysis details}

The data for this paper were collected in the photoproduction
experiment FOCUS during the Fermilab 1996--1997 fixed-target run. In
FOCUS, a forward multi-particle spectrometer is used to measure the
interactions of high energy photons on a segmented BeO target. The
FOCUS detector is a large aperture, fixed-target spectrometer with
excellent vertexing and particle identification. Most of the FOCUS
experiment and analysis techniques have been described
previously~\cite{anomaly,ycp,CNIM}. In this section we describe the cuts used
both in the non-parametric analysis described in Section 3 as well as the parametric analysis described
in Section 4.

The non-parametric part of this analysis is based on a sample of $\approx 13,000$ decays of the form \dstardk{}, where \kmunudk{}.
To isolate the \kmunudk{} topology, we required that the muon, and kaon
tracks appeared in a secondary vertex with a confidence level
exceeding 1\%.  In order to suppress backgrounds from higher multiplicity charm decays, we isolated
the $K^- \mu^+$ vertex from other tracks (not including tracks from the primary vertex) by 
requiring that the maximum confidence level for another track to form a vertex with the 
candidate be less than 1\%. The $D^*$ decay pion was required to lie in the primary vertex.

The muon track, when extrapolated to the shielded muon
arrays, was required to match muon hits with a confidence level
exceeding 1\% and all other tracks were required to have confidence level less than 1\%. 
The muon candidate was allowed
to have at most one missing hit in the 6 planes comprising our inner
muon system and a momentum exceeding 10 GeV$/c$. In order to suppress
muons from pions and kaons decaying within our apparatus, we required
that each muon candidate had a confidence level exceeding 1\% to the
hypothesis that it had a consistent trajectory through our two
analysis magnets. 
The kaon was required to have a \v Cerenkov light
pattern more consistent with that of a kaon than that of a pion by 1
unit of log likelihood \cite{CNIM}.    
Non-charm and random combinatorial 
backgrounds were reduced by requiring a detachment between the
vertex containing the \kmunudk{} and the primary production
vertex of at least 5 standard deviations.

Possible background from $D^0
\rightarrow K^- \pi^+$, where a pion is misidentified as a muon,
was reduced by requiring the reconstructed $K^- \mu^+$ mass
to be less than  $1.812~\gevcsq{}$. Finally we put a cut on the confidence level (CL$_{\rm closure}$) that the event was consistent with the hypothesis \kmunudk{} that will be described below.

\begin{figure}[tbph!]
 \begin{center}
\includegraphics[width=3.in]{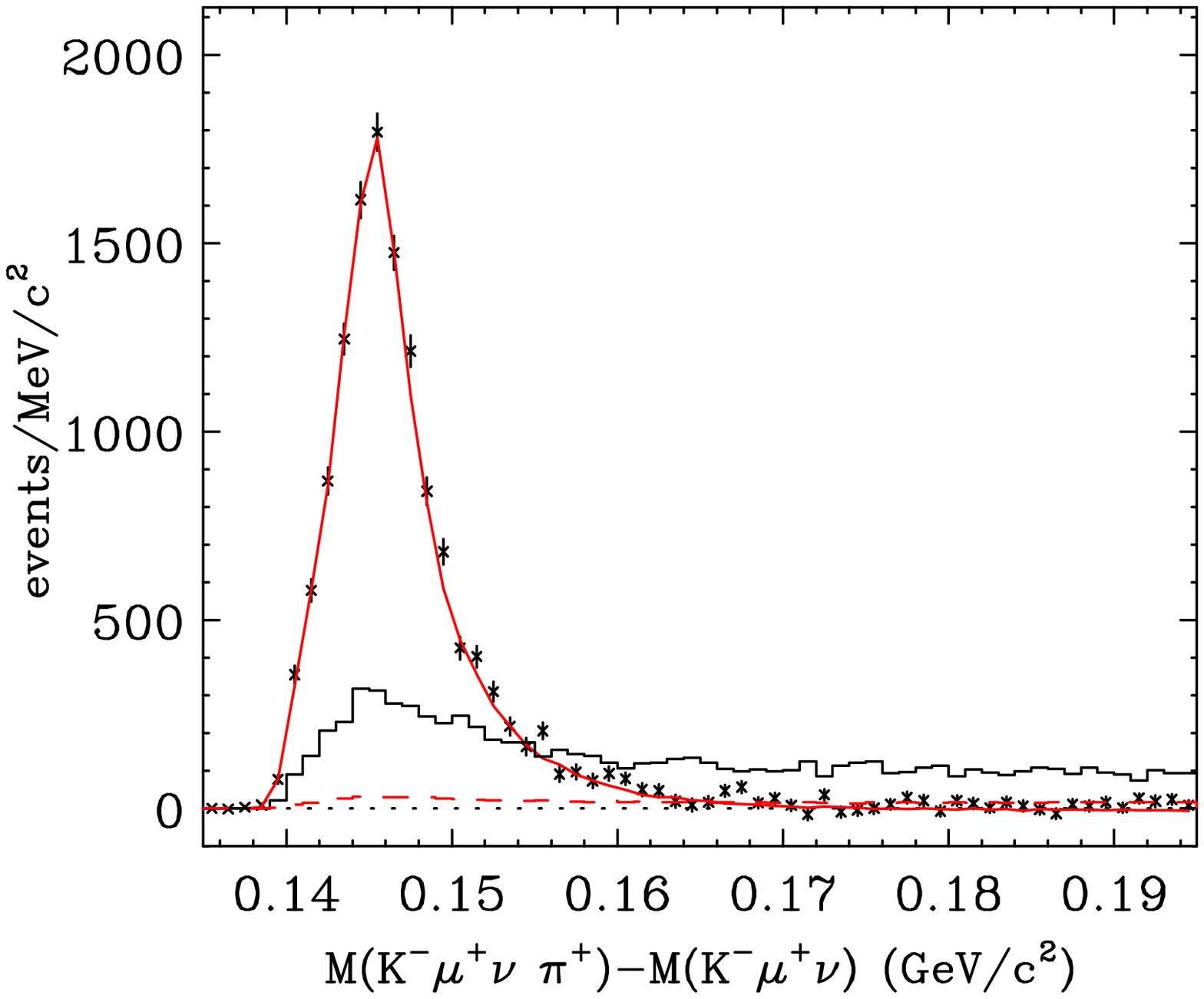}
\caption{ The $D^* - D$ mass difference distribution for events satisfying our signal selection cuts.
The points with error bars are for wrong-sign subtracted data.  The histogram shows the mass difference distribution obtained in data for 
the wrong sign events. The solid curve shows the wrong-sign subtracted \mdiff{} distribution obtained in Monte Carlo; while the dashed histogram shows the \mdiff{} distribution of wrong-sign Monte Carlo events. 
There is an excess of 12,840 opposite
charge combinations over same charge combinations where the $D^* - D$ mass difference was less than 0.160 \gevcsq{}. 
\label{mdiff}}
\end{center}
\end{figure}
 
The \mdifflong{} distribution for our tagged \kmunudk{} candidates
is shown in Figure \ref{mdiff}.  Several of the distributions shown in Figure \ref{mdiff} are \emph{wrong-sign subtracted}
meaning that combinations where the $D^*$ decay pion
have the same charge as the kaon are subtracted from those where the decay pion and kaon have the opposite 
charge. Figure \ref{mdiff} was created using our \emph{standard} \cite{focus_ff} line-of-flight neutrino closure technique. Briefly, the standard neutrino closure method assumes the reconstructed D momentum vector points along the displacement between the secondary and primary vertex. This leaves a two-fold ambiguity on the neutrino momentum. For Figure \ref{mdiff}, we use the neutrino momentum that resulted in the lower \mdiff{}.  Figure \ref{mdiff} illustrates the necessity of making a wrong-sign subtraction, since the wrong-sign 
fraction in the data is much larger than predicted by our charm Monte Carlo, indicating the 
presence of a non-negligible non-charm background in the data.  We believe that this non-charm
background is right-sign, wrong-sign symmetric.

We next describe the cuts used for the two-dimensional fit analysis described in Section 4. 
One of the principal motivations for this analysis, was to compare the decay widths for \pimunudk{} and 
\kmunudk{} and extract the ratio $f_+^{(\pi)}(0)/f_+^{(K)} (0)$. As such, the analysis cuts used for the two-dimensional analysis
are somewhat different than those previously described for the deconvolution analysis in order to reduce systematic uncertainty on the ratio.  Several additional cuts on the muon candidate were applied  to remove contamination from electrons. 
The kaon, in \kmunudk{}, was required to have a \v Cerenkov light
pattern more consistent with that for a kaon than with that for a pion by 3
units of log likelihood.  The pion track, in \pimunudk{}, was required to have a \v Cerenkov light
pattern more consistent with that of a pion than that of a kaon by 3
units of log likelihood.  For both the pion from \pimunudk{} decay as well as the $D^{*+}$ decay pion, we further required that no other hypothesis was favored over the pion hypothesis by more than 6 units of likelihood. 
The pion in \pimunudk{} was required to have a momentum greater than 14 GeV$/c$ , and the $D^*$ decay pion was required to have a momentum greater than 2.5 GeV$/c$. A $D^* - D$  mass difference cut of $\mdiff{} < 0.154~\gevcsq{}$ was applied. Finally the hadron-muon mass was required to exceed 1 \gevcsq{}.

To improve the \qsq{} resolution for both the parametric and non-parametric analyses, we developed an alternative neutrino closure
technique that we will call $D^*$ \emph{cone closure}. We require that the
$K^- \mu^+ \nu$ reconstructs to the mass of a $D^0$ and the  $K^- \mu^+ \nu ~\pi^+$ reconstructs
to the mass of $D^{*+}$.  When viewed in the $K^- \mu^+$ rest frame, these constraints place the neutrino momentum vector
on a cone about the $D^*$ decay pion where both the neutrino energy and cone half-angle are determined from 
the mass constraints and the well measured $K^-$, $\mu^+$, and $\pi^+$ momentum vectors.  We then sample all azimuths for the neutrino in this cone, reconstruct the lab frame $D^0$ momentum vector, and choose the azimuth where the $D^0$ is most consistent with pointing to the primary vertex based on minimizing a $\chi^2$ variable. In order to further reduce backgrounds, we required CL$_{\rm closure} > 1\%$ where CL$_{\rm closure}$ is a confidence level based this minimal $\chi^2$. 

\begin{figure}[tbph!]
 \begin{center}
\includegraphics[width=3.in]{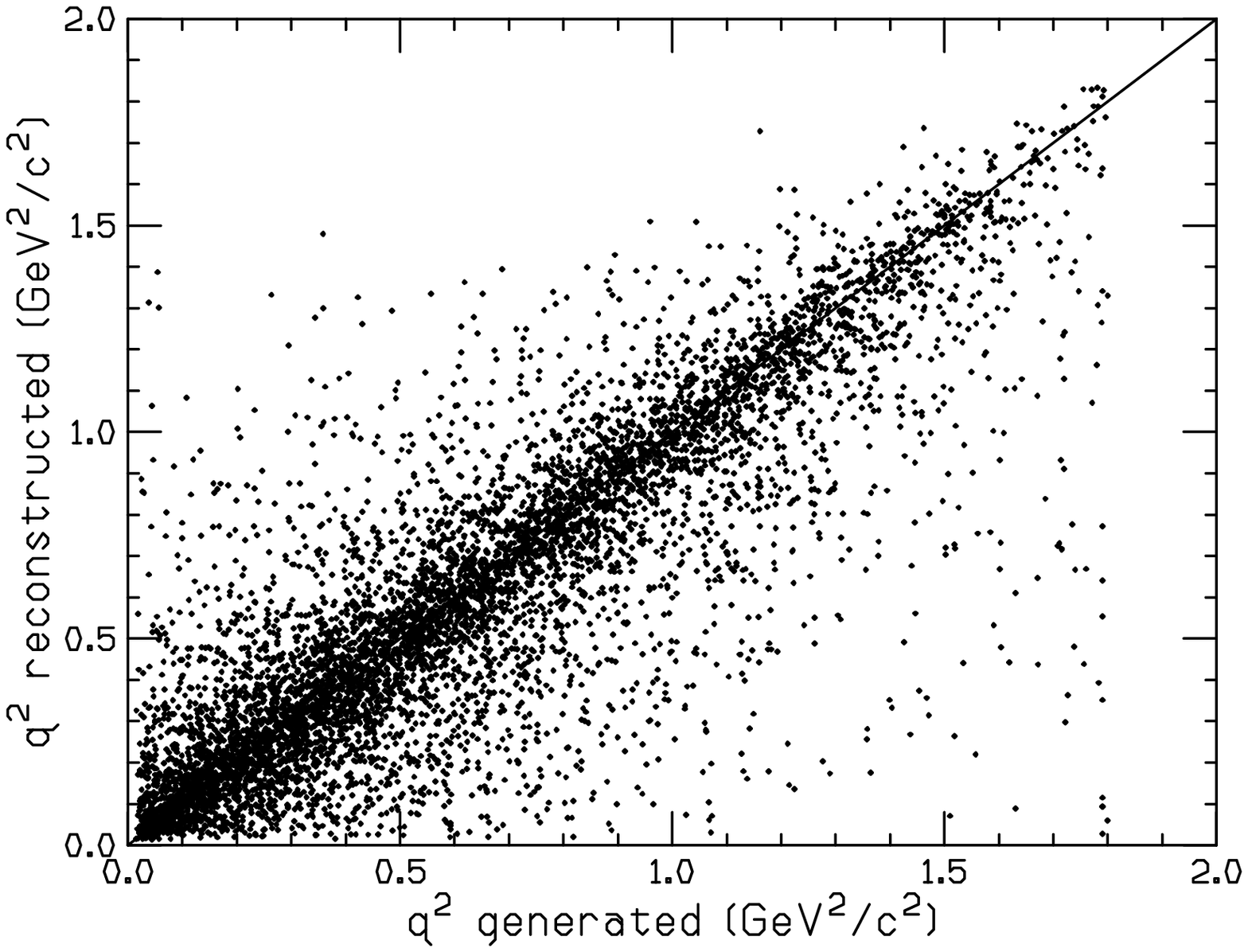}
\caption{ A scatter plot of \qsq{} reconstructed by the cone closure technique and the true, generated \qsq{} in Monte Carlo events.  We show the line where \qsq{} generated equals 
\qsq{} reconstructed. 
\label{qsqRECvGEN}}
\end{center}
\end{figure}

Averaged over all detected \kmunudk{} events, the Monte Carlo predicted a rather
non-Gaussian \qsq{} resolution with an r.m.s. width of 0.22 \gevc{} using the $D^*$ cone closure technique.  The correlation between the generated and reconstructed \qsq{} is illustrated in Figure \ref{qsqRECvGEN}.

It was important to test the fidelity of the simulation with respect
to the reproducibility of the \qsq{} resolution.  To do this, we studied tagged, 
fully-reconstructed $D^0 \rightarrow K^- \pi^+ \pi^+ \pi^+$ decays from \dstardk{}
where, as a test, one of the $D^0$ decay pions was reconstructed using our
neutrino cone closure technique. We then reconstructed the \qsq{}
using the neutrino closure and compared it to a precisely reconstructed \qsq{} obtained
from the magnetically reconstructed ``neutrino" pion. The difference between 
these two \qsq{} values provided a resolution distribution obtained from data that
could then be compared to the same resolution distribution obtained using tagged 
$D^0 \rightarrow K^- \pi^+ \pi^+ \pi^+$ in our Monte Carlo. The Monte Carlo resolution distribution
was a good match to the observed resolution distribution.

\mysection{Non-parametric analysis of \fplusK{}}
We begin with a description of the method used to correct for the effects of acceptance
and \qsq{} resolution. We will call this our \emph{deconvolution} technique. The goal of the
deconvolution is to produce a set of \fplusbinsq{} that represents measured
\fplusK{} values - each averaged over the (narrow) width of the $q^2_i$ \emph{reporting} bins. 
Under the assumption that $\fminus /\fplus$ is on the order of unity, Eq. \ref{dgamdq} implies that the number of events
expected in a given \qsq{} bin is proportional to $|\fplus|^2$.  Our Monte Carlo is used
to determine the fraction of events reconstructed in a given \qsq{} bin that were 
generated in another \qsq{} bin. This information, along with the \fplusK{} distribution used
in the original generation\footnote{The sample was generated
assuming $f_-(\qsq)/f_+(\qsq) = -0.7$} of the \kmunudk{} Monte Carlo sample, was combined to form a square matrix that linearly relates a vector of the predicted number of events reconstructed in each \qsq{} bin to a column vector of assumed \fplusbinsq{} values. 

\begin{eqnarray}
 \pmatrix{n_1 \cr n_2 \cr \vdots} = 
\eta~\pmatrix{ {N_{11} / \tilde f^2_+ (q^2_1)} & {N_{12} / \tilde f^2_+ (q^2_2)} & \ldots \cr
{N_{21} / \tilde f^2_+ (q^2_1)} & {N_{22} / \tilde f^2_+ (q^2_2)} & \ldots \cr
\vdots & \ldots & \ddots} \pmatrix{f^2_+ (q^2_1)\cr  f^2_+ (q^2_2) \cr \vdots}
 \label{deconv}
\end{eqnarray}
In Eq. \ref{deconv}, $n_i$ are the number of events that are observed in the $i$th \qsq{} bin, $\eta$ is the
ratio of the observed to the number of generated events,
$N_{i j}$ is the number of Monte Carlo events that were generated in \qsq{} bin $j$ that reconstruct
in \qsq{} bin $i$, $\tilde f^2_+ (q^2_i)$ are the input $f^2_+ (q^2)$ used in the Monte Carlo generation, and \fplusbinsq{} are
the true form factors that describe the data. 

With reference to Eq. \ref{deconv}, the ``deconvolved" \fplusbinsq{} is then given by the inverse of the square matrix times the first column vector that consists of the observed number of events reconstructed in each \qsq{} bin. 
We will call the inverse of this matrix the \emph{deconvolution} matrix with components $D_{i j}$. 
In this notation: $\fplusbinsq{} = \sum_j~D_{i j}~n_j$. 

\begin{figure}[tbph!]
 \begin{center}
\includegraphics[width=3.3in]{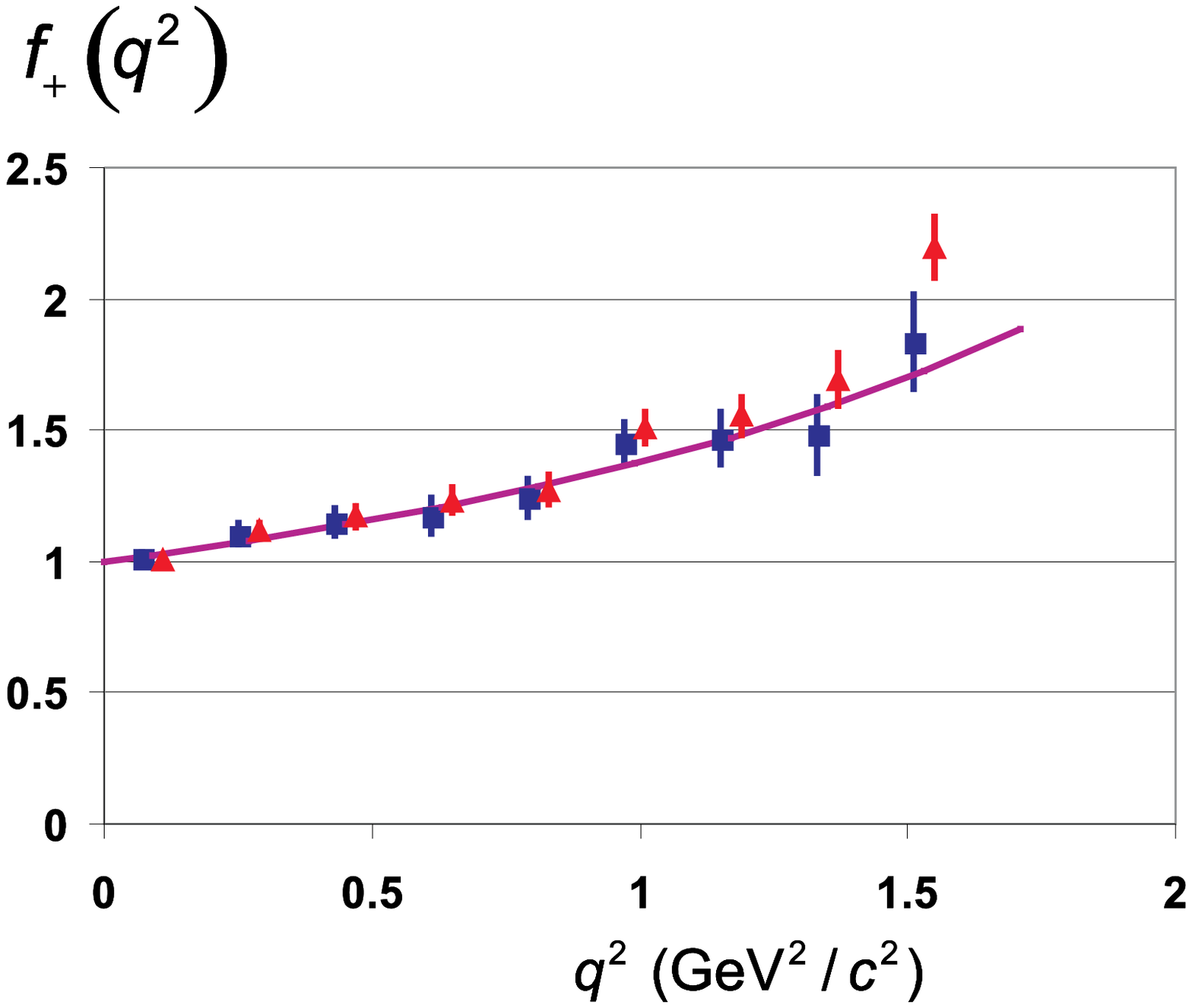}
\caption{The deconvolved \fplusK{} for \kmunudk{} events using nine, 0.18 \gevc{} bins. The triangular points
are before the known charm backgrounds were subtracted. The square points are after subtraction for known
charm backgrounds. The line shows the pole form with \mpole{} = 1.91~\gevcsq{} -- the value obtained
from a fit to the displayed \fplusK{} points. After background subtraction, the confidence level of the fit to the pole form is 87\%.
A fit to the modified pole form produced an $\alpha$ parameter of $0.32$ with a confidence level of 82\%.
\label{subtract}}
\end{center}
\end{figure}

We performed the sum $\fplusbinsq{} = \sum_j~D_{i j} n_j$ by using a  
separate, weighted histogram for each \fplusbinsq{}.  Each \fplusbinsq{} is a sum
of weights over all events where the event weight $D_{i j}$ where $i$ is $q^2_i$ \emph{reporting} bin and $j$ is the \emph{reconstructed} $q^2_j$ bin for that event. We perform
a wrong-sign subtraction by multiplying the deconvolution weight $D_{ij}$ by +1 if the kaon of the event had the opposite sign of the $D^*$ decay pion and -1 otherwise. 
Our charm background correction is based on a Monte Carlo, which incorporates all known charm decays and charm decay mechanisms. The charm background was normalized to the same number of \dstardk{},
\kpidk{} events observed in the data. We subtract known backgrounds by deducting the deconvolution weights for the background events predicted by our charm Monte Carlo. 
Figure \ref{subtract} compares the \fplusbin{} values
obtained with and without the background subtraction.\footnote{
The covariance between two \qsq{} reporting bins is given by the sum of the product of event weights for the two reporting bins. We take the square root of the \fplusbinsq{} returned by the fit and make the appropriate adjustment to the variances obtained from the diagonal elements of the covariance matrix.}   Figure \ref{subtract} shows that
most of the charm background is expected in the high \qsq{} region, and once the background
is subtracted, the data is an excellent fit to the pole form.
In the range $\mdiff < 0. 16~\gevcsq{}$, the
expected, wrong-sign subtracted background yield from our Monte Carlo was found to be 12.6 \% of the total 
number of events in this \mdiff{} range when using our baseline cuts. 

The deconvolution was obtained by summing the weights of all events with $\mdiff < 0.16~\gevcsq{}$.
A ten bin deconvolution matrix was used, with the overflow bin dropped. 
Given that our bin width, 0.18 \gevc{}, is comparable to our r.m.s. \qsq{} resolution, adjacent
\fplusbin{} values have a strong, negative correlation (typically - 65\%) and the error bars
are thus significantly inflated over naive counting statistics errors.  
Table \ref{fqsq_table} summarizes of our non-parametric \fplusK{} measurements for
\kmunudk{} along with the correlation matrix.  The \fplusK{} values are normalized such that $f_+(0) = 1$.
\begin{table}[t]
\centering
\caption{Measurements of \fplusK{} for \kmunudk{} along with the 9 $\times$ 9
matrix of relative correlation coefficients.
}
\vskip 0.1 in
\begin{tabular}{|l|c|c|c|c|c|c|c|c|c|c|c|c|c|}
\hline
 $i$ & $q^2_i$ & $f_+(q^2_i)$& &    &   1  &   2   &   3  &   4   & 5    &  6   &  7   & 8 & 9 \\
\hline
1 & 0.09 & 1.01 $\pm$ 0.03 & &  1 & 1.00 & -0.63 & 0.25 & -0.10 & 0.03 & -0.01 & 0.00 & 0.00 & 0.00 \\ 
2 & 0.27 & 1.11 $\pm$ 0.05 & &  2 & -0.63 & 1.00 & -0.68 & 0.29 & -0.11 & 0.04 & -0.01 & 0.01 & -0.01 \\ 
3 & 0.45 & 1.15 $\pm$ 0.07 & & 3 & 0.25 & -0.68 & 1.00 & -0.68 & 0.27 & -0.10 & 0.03 & -0.01 & 0.01 \\ 
4 & 0.63 & 1.17 $\pm$ 0.08 & & 4 & -0.10 & 0.29 & -0.68 & 1.00 & -0.65 & 0.26 & -0.09 & 0.03 & -0.02 \\ 
5 & 0.81 & 1.24 $\pm$ 0.09 & & 5 & 0.03 & -0.11 & 0.27 & -0.65 & 1.00 & -0.65 & 0.23 & -0.08 & 0.03 \\ 
6 & 0.99 & 1.45 $\pm$ 0.09 & & 6 & -0.01 & 0.04 & -0.10 & 0.26 & -0.65 & 1.00 & -0.60 & 0.20 & -0.07 \\ 
7 & 1.17 & 1.47 $\pm$ 0.11 & & 7 & 0.00 & -0.01 & 0.03 & -0.09 & 0.23 & -0.60 & 1.00 & -0.58 & 0.19 \\ 
8 & 1.35 & 1.48 $\pm$ 0.16 & & 8 & 0.00 & 0.01 & -0.01 & 0.03 & -0.08 & 0.20 & -0.58 & 1.00 & -0.56 \\ 
9 & 1.53 & 1.84 $\pm$ 0.19 & & 9 & 0.00 & -0.01 & 0.01 & -0.02 & 0.03 & -0.07 & 0.19 & -0.56 & 1.00 \\
\hline
\end{tabular}
\label{fqsq_table}
\end{table}
\clearpage

\section{Parameterized \fplus{} forms for \kmunudk{} and \pimunudk{}}

In this section we present values of the \mpole{} and $\alpha$ parameters for the pole form (Eq. \ref{pole}) 
and modified pole form (Eq. \ref{modpole}) as well as the ratio \etak{}.  For the case of \kmunudk{} we have done this directly from 
a $\chi^2$ fit of the non-parametric \fplusK{} values illustrated in Figure \ref{subtract} as well as from a two-dimensional binned likelihood fit to the \qsq{} and \costhl{} scatterplot where \thl{} is the angle between
the $\nu$ and the $D$ direction in the $\mu\nu$ rest frame.  Because of the much larger background contamination in 
the Cabibbo suppressed \pimunudk{}, only 
the two-dimensional binned likelihood fit was employed to extract \fplusK{} parameters for this mode.  We begin with a discussion
of the results from the two-dimensional fit.

Information on parameterized \fplusK{} and \etak{} is obtained by using a weighting technique that is similar to that described
in \cite{focus_ff}. We use a binned version of the fitting technique developed
by the E691 collaboration~\cite{schmidt} for fitting decay intensities
where the kinematic variables that rely on reconstructed neutrino kinematics are poorly measured.
The observed number of events in each \qsq{}-\costhl{} 
bin is compared to a prediction based on signal intensity as well as background contributions. 

The signal component is constructed from a weighted Monte Carlo. 
The signal Monte Carlo was initially generated using nominal values for $f_-(0)/f_+(0)$ and \mpole{}. Both the generated
as well as the reconstructed kinematic variables were stored for each
event. The signal prediction for a given fit iteration is then
computed by weighting each event within a given reconstructed
kinematic bin by the intensity evaluated
using the generated kinematic variables for the current set of fit
parameters divided by the generated intensity.  
 
A variety of possible backgrounds were included for our two processes.
These included general charm background based on our charm Monte Carlo as well as specific backgrounds
that create peaks in the \mdiff{} distribution.
For the case of \pimunudk{}, the specific backgrounds included $K^- \mu^+ \nu$, $K^- \pi^0 \mu^+ \nu$, $\overline{K}^0 \pi^- \mu^+ \nu$
and $\rho^- \mu \nu$ ; while for \kmunudk{} this included $K^- \pi^0 \mu^+ \nu$.  
In all cases, the shape of the backgrounds were determined from our Monte Carlo that incorporated known decay intensities. The branching ratio of each specific background relative to the two signal processes were allowed to float, but a $\chi^2$ (likelihood penalty term) was included to tie a given background's branching ratio relative to the signal to the measured values within their known uncertainties. The yield of \kmunudk{} deduced from a fit to its \qsq{}-\costhl{} scatterplot served as an estimate of this important background in the fit to the \qsq{}-\costhl{} scatterplot for \pimunudk{}.  A more complete description of this fitting procedure along with our value for $f_+^{(\pi)}(0)/f_+^{(K)} (0)$
appears in a companion paper \cite{companion}. 
\begin{figure}[htp]
\begin{center}
\includegraphics[width=3.5in]{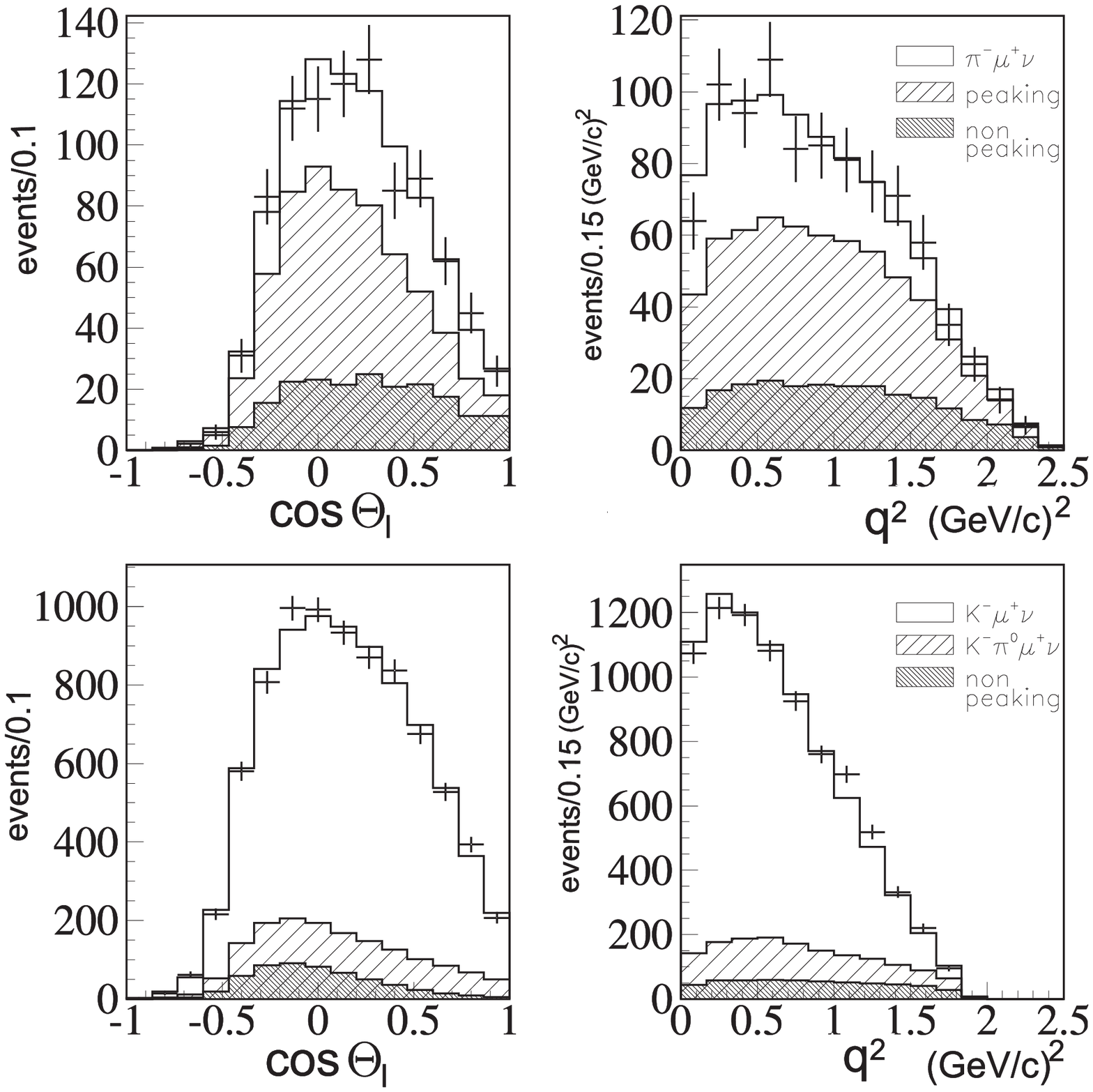}
\end{center}
\caption{
The \qsq{} and \costhl{} projections of the two-dimensional fit compared to the data histogram for both \pimunudk{} (upper row) and \kmunudk{} (lower row). For the case of \pimunudk{}, the peaking backgrounds included the sum of $K^- \mu^+ \nu$, $K^- \pi^0 \mu^+ \nu$, $\overline{K}^0 \pi^- \mu^+ \nu$
and $\rho^- \mu \nu$ ; while for \kmunudk{} they included $K^- \pi^0 \mu^+ \nu$.  
\label{composit}}
\end{figure}

Figure \ref{composit} shows how the \qsq{} and \costhl{} projections predicted by the fit compare to the data as well as the various signal and background components of these projections. 
The results relevant to  \kmunudk{} are \mpoleresult{}, \alpharesult{}, and \etakresult{}.

The systematic error was determined by comparing results using different event selections, alternative fit methods, and looking at the consistency of results between split samples.  
We begin with some of the many alternative event selections that were investigated. 
A \fplusK{} and \etak{} measurement was
obtained from fits where each of these cuts was varied relative to our baseline: the detachment of the secondary vertex from the primary vertex was varied from 4 to 12 standard deviations, the secondary vertex was required to lie out of all target material, the momentum cut on the muon was raised from 10 to 25 GeV$/c$, the secondary isolation cut was tightened
from $< 1 \%$ to $< 0.1 \%$ and the confidence level on the secondary vertex was raised from $> 1\%$ to $> 15\%$.  The split sample compared the form factor information for particles to that for antiparticles.  Various alternative fits
were employed. For example, in some fits, the two pole masses were allowed to float while keeping \etak{} fixed compared
to our standard fit where all three parameters were free to float. In another fit variant, the fit was performed
on the \mdiff{} - \qsq{} scatter plot as opposed to the \qsq{} - \costhl{} scatterplot. 

Finally a fit to \mpole{} was made directly from the non-parametric \fplusK{} results illustrated in Figure \ref{subtract}.  This fit minimized a $\chi^2$ given by given by Eq. \ref{chisq}.
\begin{eqnarray}
 \chi^2 =& \sum_i \sum_j \left(\fplusbin^{(m)} - \fplusbin^{(p)} \right)~C^{-1}_{ij} 
\left(\fplusbinJ^{(m)} - \fplusbinJ^{(p)} \right) \cr
& + \left(b - 1 \over \sigma_b \right)^2
 \label{chisq}
\end{eqnarray}
where the sum runs over all reporting bins, $C^{-1}$ is the inverse of the covariance matrix,  $\fplusbin^{(m)}$ are the measured \fplusK{} values, and 
$\fplusbin^{(p)}$ are the predicted \fplusK{} within the parameterization. The second term is 
a likelihood penalty term that parameterizes uncertainty in the level of the charm background. The parameter $b$ is a background multiplier that multiplies the expected Monte Carlo background yield and $\sigma_b$ is our estimate of its uncertainty.
The parameterized $\fplusbin^{(p)}$ depends on a normalization parameter $f_+(0)$
and a shape parameter \mpole{}.  This fit returned a pole mass of \mpoleresultdecon{} which is in remarkably
good agreement with \mpoleresult{}--the value obtained from the two-dimensional, parametric fit. Again, the systematic error of the fit to the non-parametric \fplusK{} was
obtained by checking its stability against a variety of different fit variants, cut variants, assumed background
levels, and $f_-/f_+$ assumptions.
 
Using a procedure identical to the parametric procedure used for \kmunudk{}, we find \mpole{} value for \pimunudk{} is \mpoleresultpi{}.  The systematic error on this result
included an additional important cut variant consisting of raising the log-likelihood difference between the kaon and pion Cerenkov hypothesis from 3 to 5 and in the process reducing the fraction of kaons misidentified as pions by about a factor of two. 

\mysection{Summary}

\begin{figure}[htp]
\begin{center}
\includegraphics[width=4.in]{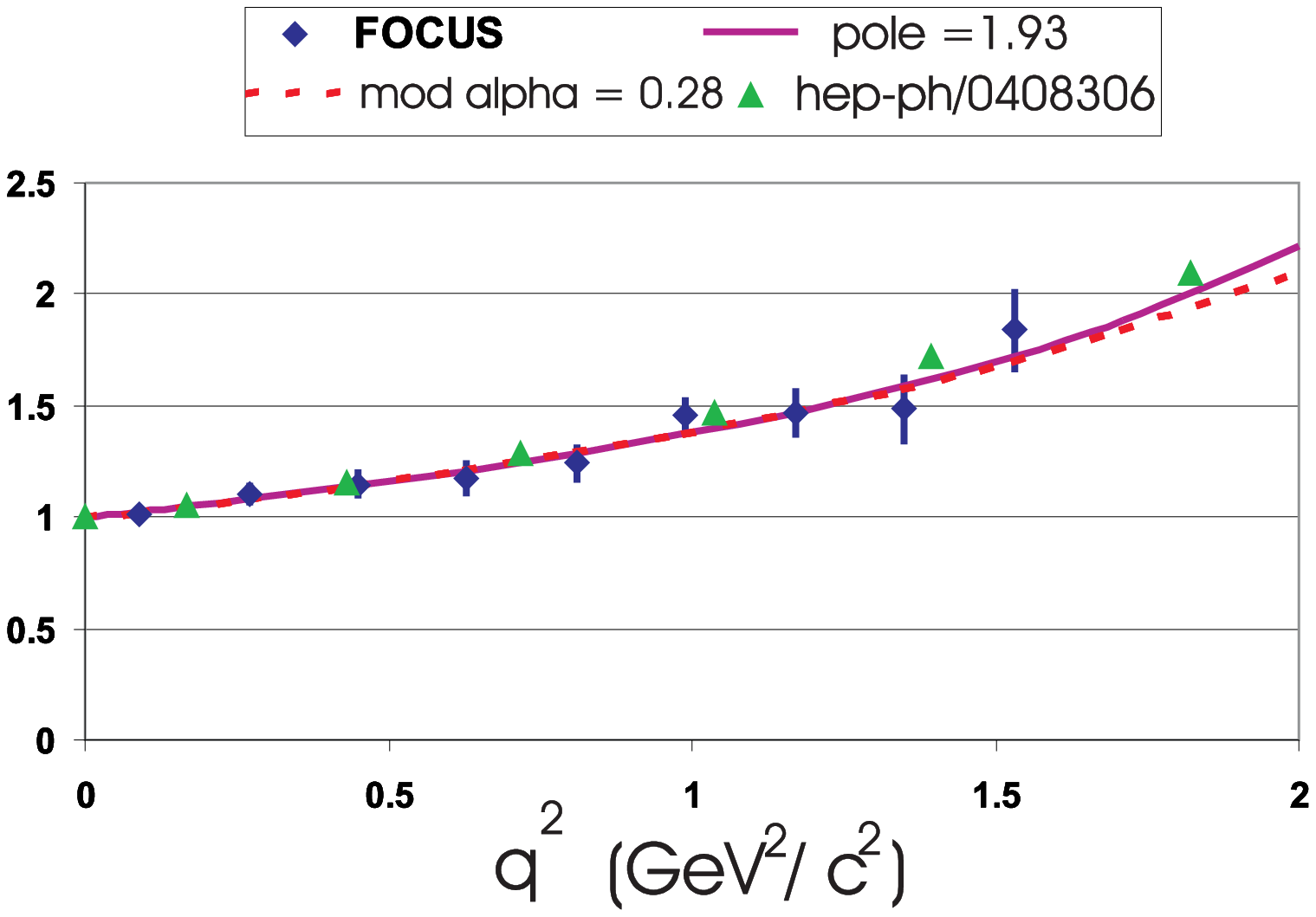}
\end{center}
\caption{
The background subtracted \fplusK{} (diamonds with error bars) is compared to a pole form with \mpole{} = 1.93 \gevcsq{} (solid
curve) , a modified pole form with $\alpha = 0.28$ (dashed curve) , and the
unquenched, Lattice QCD, calculations given in reference \cite{simone} (triangles with no error bars).  
This form factor is for the process \kmunudk{}. The $\alpha$ and \mpole{} used for the plots are 
obtained using the two-dimensional, parameterized fit.  
\label{fqsq_kmunu}}
\end{figure}

Figure \ref{fqsq_kmunu} compares our \fplusbin{} measurements to a recent \cite{simone} Lattice QCD calculation\footnote{We re-scaled their calculations to insure that $f_+(0) = 1$.} and
our best fit values for \mpole{} in the pole mass parameterization (Eq. \ref{pole}) and
$\alpha$ in the modified pole mass parameterization (Eq. \ref{modpole}).
We obtained a value of \etakresult{} that is also consistent with the value that can be derived from information in \cite{simone}.
A tabular summary of the data of Figure \ref{fqsq_kmunu} and its correlation matrix
has appeared in Table 1 of Section 3.

\begin{figure}[htp]
\begin{center}
\includegraphics[width=4.5in]{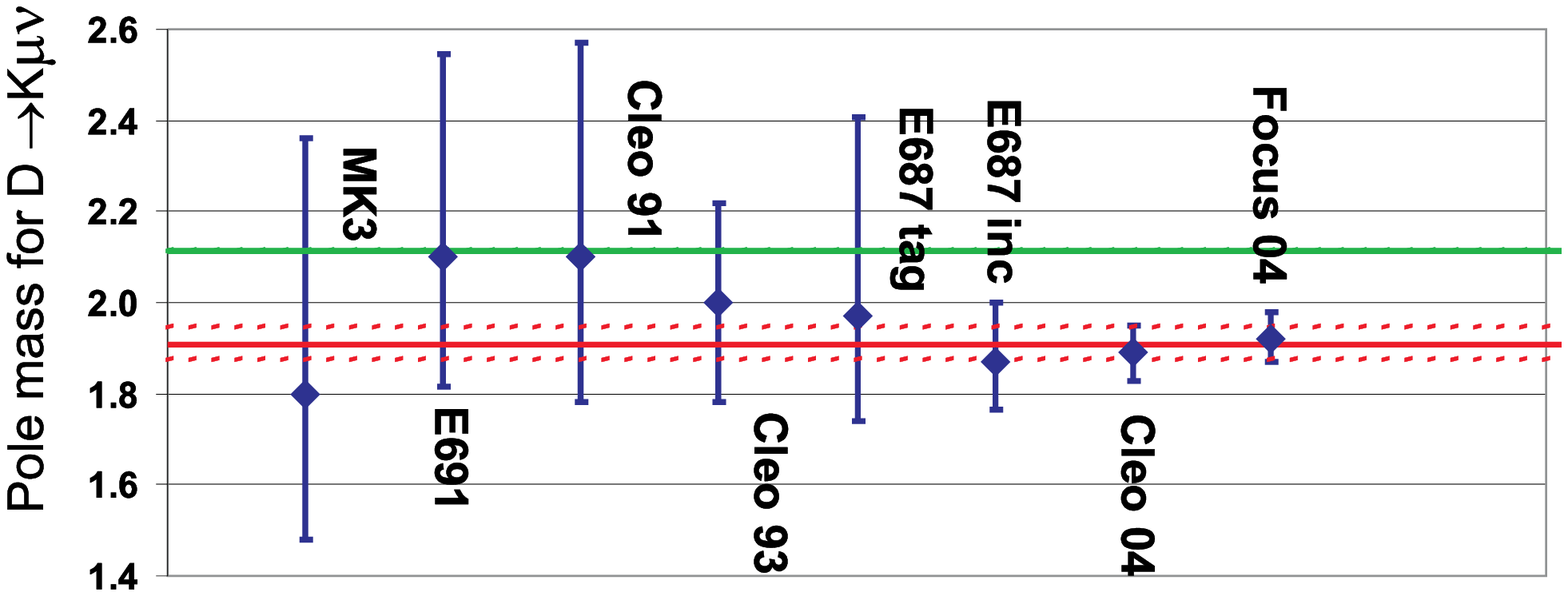}
\end{center}
\caption{Summary of \mpole{} measurements. All data are consistent with a weighted average pole mass
of $\mpole = 1.91 \pm 0.04~\gevcsq$. The upper solid line shows the spectroscopic pole mass at \mdsstar{}. The lower solid line and two dashed lines represent the weighted average and its error.  
Our weighted average of all data is 5.1 $\sigma$ lower than \mdsstar{}.
\label{kmunupole}}
\end{figure}

Our fit to the \mpole{} parameter in pole mass parameterization was \mpoleresult{}.  This is compared
to previous published data in Figure \ref{kmunupole}. The most recent \mpole{} is from CLEO \cite{cleopilnu} who obtain 
$\mpole = 1.89 \pm 0.05 \pm 0.035~\gevcsq$.
All data are remarkably consistent.

Our fit to the $\alpha$ parameter for the modified pole form is \alpharesult{} from the parameterized, two-dimensional fit. 
This is very consistent
with 0.32 $\pm$ 0.09 $\pm$ 0.07 , the value obtained from our fits to non-parametric data shown in Figure \ref{subtract}. 
The most recent
published measurement is from CLEO \cite{cleopilnu} who obtain $\alpha = 0.36 \pm 0.10^{+0.03}_{-0.07}$.
Our value for the $\alpha$ parameter is 1.9 $\sigma$ lower than the value quoted in \cite{simone} for \kmunudk{}.\footnote{We believe that only statistical
errors on $\alpha$ are included in \cite{simone}}

We also find that \mpole{} for \pimunudk{} is \mpoleresultpi{}. This value is compatible
with our value for the pole mass for \kmunudk{}. In the naive pole dominance model, 
the \mpole{} for \pimunudk{} would be at the mass of the $D^{*+}$ and would therefore
lie lower in mass than \mdsstar{} expected for \kmunudk{}.


\mysection{Acknowledgments}

We wish to acknowledge the assistance of the staffs of Fermi National
Accelerator Laboratory, the INFN of Italy, and the physics departments
of the collaborating institutions. This research was supported in part
by the U.~S.  National Science Foundation, the U.~S. Department of
Energy, the Italian Istituto Nazionale di Fisica Nucleare and
Ministero dell'Istruzione dell'Universit\`a e della Ricerca, the
Brazilian Conselho Nacional de Desenvolvimento Cient\'{\i}fico e
Tecnol\'ogico, CONACyT-M\'exico, the Korean Ministry of Education, 
and the Korean Science and Engineering Foundation.

\end{document}

%% file: authors_revised.tex
\collaboration{The FOCUS Collaboration}

\author[ucd]{J.~M.~Link}
\author[ucd]{P.~M.~Yager}
\author[cbpf]{J.~C.~Anjos}
\author[cbpf]{I.~Bediaga}
\author[cbpf]{C.~G\"obel}
\author[cbpf]{A.~A.~Machado}
\author[cbpf]{J.~Magnin}
\author[cbpf]{A.~Massafferri}
\author[cbpf]{J.~M.~de~Miranda}
\author[cbpf]{I.~M.~Pepe}
% ...

\address[ucd]{University of California, Davis, CA 95616}
\address[cbpf]{Centro Brasileiro de Pesquisas F\'{\i}sicas, Rio de Janeiro, RJ, Brasil}
% ...
